\documentclass[lettersize,journal]{IEEEtran}
\usepackage{amsmath,amssymb,amsfonts}
\usepackage{algorithmic}
\usepackage{algorithm}
\usepackage{array}
\usepackage[caption=false,font=footnotesize,labelfont=rm,textfont=rm]{subfig}
\usepackage{textcomp}
\usepackage{stfloats}
\usepackage{url}
\usepackage{verbatim}
\usepackage{graphicx}
\usepackage{soul}
\usepackage{siunitx}
\usepackage{cite}
\usepackage{tabularx} 
\usepackage{xcolor}
\usepackage{threeparttable} 
\usepackage{booktabs}
\usepackage{multirow} 
\usepackage{comment}

\begin{document}

\title{Adaptive Uncertainty-Aware Modeling and Stochastic Radial Basis Function Predictive Control for Personalized Fluid Resuscitation}

\author{Elham Estiri, \IEEEmembership{Student Member, IEEE}, Hossein Mirinejad, \IEEEmembership{Senior Member, IEEE}
\thanks{This work was supported by the National Science 
Foundation CAREER Award under Grant No. 2340139 and by the National 
Science Foundation Engineering Research Initiation (ERI) Award under Grant 
No. 2138929.}
\thanks{Elham Estiri is with the School of Engineering, College of Aeronautics and Engineering, Kent State University, Kent, OH 44242 USA (e-mail: eestiri@kent.edu).}
\thanks{Hossein Mirinejad is with the School of Engineering, College of Aeronautics and Engineering, Kent State University, Kent, OH 44242 USA (correspondence e-mail: hmiri@kent.edu).}
}

\markboth{Journal of \LaTeX\ Class Files,~Vol.~14, No.~8, August~2021}%
{Shell \MakeLowercase{\textit{et al.}}: A Sample Article Using IEEEtran.cls for IEEE Journals}


\maketitle

\begin{abstract}
This paper presents a novel framework integrating Bayesian physiological modeling with optimal control strategies to achieve uncertainty-aware, personalized hemodynamic regulation during fluid resuscitation. An uncertainty-aware variational autoencoder state-space model (UVAE-SSM) was first developed to capture the dynamical relationship between mean arterial pressure (MAP) and fluid infusion using limited data, while explicitly modeling aleatoric uncertainty (i.e., randomness in the measurements, such as sensor noise). Then, a Bayesian nonlinear state-space model (BNSSM) was developed by utilizing Bayesian neural networks (BNNs) to capture epistemic uncertainty arising from physiological and patient-specific variability, enabling the creation of a virtual patient generator (VPG). Building on this uncertainty-aware modeling framework, a stochastic radial basis function model predictive control (sRBF-MPC) algorithm was designed to track the MAP target while satisfying physiological constraints. Finally, an online fine-tuning algorithm was developed to adapt the nominal UVAE-SSM  using streaming VPG data, enabling progressive personalization during closed-loop therapy. Simulation results across unseen animal subjects and an independent human clinical dataset demonstrated the strong predictive accuracy and cross-population generalizability of the UVAE-SSM and BNSSM models. Closed-loop evaluations confirmed that the proposed sRBF-MPC framework achieved stable MAP regulation while providing better risk-aware control compared to quadratic MPC (Q-MPC) and  stochastic quadratic MPC (sQ-MPC). Overall, the proposed framework accounts for inter- and intra-patient variability through online model adaptation, offering a promising step toward uncertainty-aware, personalized hemodynamic modeling and control in critical care. 
\end{abstract}


\section{Introduction}

\IEEEPARstart{H}{emorrhage} is a critical condition in trauma and perioperative care where uncontrolled blood loss leads to rapid deterioration of cardiovascular stability \cite{avital2022closed}. Timely fluid resuscitation is the first-line intervention, but determining the optimal infusion rate is highly challenging due to large inter- and intra-patient variability in hemodynamic responses. Clinicians often rely on standardized protocols such as advanced life support (ALS), recommending fixed-rate infusions without accounting for individual patient characteristics \cite{mohammad2014educational}. This lack of personalization can result in underdosing, where insufficient fluids fail to adequately restore perfusion, or overdosing, where excess fluids exacerbate hemodilution, coagulopathy, or tissue edema \cite{parvinian2019credibility}. Closed-loop fluid resuscitation systems have been proposed to mitigate these risks by continuously adjusting infusion rates in response to patient feedback, yet existing approaches face limitations in both patient modeling and controller design \cite{estiri2023precision}.

Prior modeling approaches for closed-loop resuscitation have largely relied on simplified or lumped-parameter formulations \cite{bighamian2016lumped, bighamian2018control}. While suitable for controller design, these models lack fidelity in capturing nonlinear hemodynamics and subject-specific variability. Additionally, existing modeling frameworks largely neglect clinical uncertainty, limiting their clinical applicability. In practice, clinical data are often sparse and distorted by various sources of uncertainty such as measurement noise, underscoring the need for models that explicitly account for aleatoric uncertainty, i.e., the irreducible uncertainty arising from noisy or imperfect observations\cite{kim2014accuracy}. 

Existing simulation-based resuscitation studies have largely employed the same mathematical model to represent the patient and to design the controller \cite{grant2023optimal, jin2018development, meerza2021precise, alsalti2022design}. This assumption is clinically unrealistic since the controller does not have direct access to true patient dynamics. In frameworks such as model predictive control (MPC), treatment decisions depend on forecasting patient responses to candidate infusions \cite{estiri2024variational}. This dependence underscores the need for a distinct predictive surrogate that can imitate patient behavior during design and evaluation. A verified virtual patient generator (VPG) can fulfill this role by providing a more realistic model-based environment. Precedents for VPGs in other medical domains, including vasopressor therapy simulators \cite{kao2025development} and type 1 diabetes simulators \cite{cobelli2023developing, man2014uva}, demonstrated the clinical and regulatory potential of such platforms. However, these VPGs remain narrow in scope and none incorporate explicit quantification of epistemic uncertainty, which arises from incomplete knowledge of patient physiology and limited available data. Consequently, there remains a pressing need for an uncertainty-aware VPG for fluid resuscitation that serves as a virtual environment for autonomous systems, distinctly separated from the nominal model accessible to the controller, and capable of supporting credible evaluation under clinically realistic conditions.

Inter-patient variability in hemodynamic responses to fluid therapy is substantial, reflecting differences in physiology, injury severity, and treatment history \cite{bighamian2018control, meerza2021precise}. Intra-patient variability further complicates resuscitation, as an individual’s physiological responses evolve dynamically over the course of treatment and may be influenced by concurrent interventions or other treatments administered in intensive care. These sources of variability underscore the importance of personalization for clinically credible fluid resuscitation strategies. Existing closed-loop systems, however, typically achieve personalization in an offline manner, where models are calibrated before treatment begins and then remain fixed throughout therapy \cite{estiri2023precision, coulibaly2023personalized}. Personalized controllers based on proportional–integral-derivative (PID) \cite{mirinejad2019evaluation} or fuzzy logic \cite{rafie2004hypotensive} have demonstrated feasibility but remain inherently static, unable to adapt to evolving patient responses \cite{avital2022closed}. Broader closed-loop critical care prototypes highlight the promise of automation, yet none provide mechanisms for real-time personalization, limiting their ability to replicate the realities of clinical practice \cite{estiri2023closed, pinsky2024autonomous, ganapathy2023precision}. In a prior work, we designed a personalized fluid administration controller using a quadratic MPC (Q-MPC) algorithm \cite{estiri2024variational}. While effective in simulation, Q-MPC requires dense discretization and long prediction horizon to capture nonlinear dynamics, making real-time optimization computationally intractable. Radial basis function (RBF) methods approximate nonlinear trajectories using a compact set of basis functions, reducing the number of optimization variables and providing a tractable framework for solving nonlinear optimal control problems \cite{mirinejad2021radial, mirinejad2017individualized, estiri2026automated, mirinejad2015individualized}. However, their application to uncertainty-aware closed-loop frameworks remains unexplored.

To address the aforementioned gaps, this work develops an integrated modeling, control, and adaptation framework for automated fluid resuscitation. An uncertainty-aware variational autoencoder state-space model (UVAE-SSM) is first introduced as a nominal patient representation that learns mean arterial pressure (MAP) dynamics while capturing aleatoric uncertainty. A Bayesian nonlinear state-space model (BNSSM) is then developed using  Bayesian neural networks (BNNs) to construct a VPG that accounts for epistemic uncertainty for closed-loop evaluation. 
Building on these models, a stochastic radial basis function model predictive control (sRBF-MPC) framework is designed to propagate predictive uncertainty into the control decision, and the nominal model is fine-tuned online to personalize therapy as patient responses evolve. The main contributions are summarized as follows:

\begin{enumerate}

\item \textit{Model separation}: The controller operates on the nominal UVAE-SSM and does not assume access to true patient dynamics, which are represented independently through the BNSSM-based VPG.

\item \textit{Structured uncertainty-aware modeling and control}: Aleatoric uncertainty from the UVAE-SSM is propagated through a chance-constrained control framework, enabling uncertainty-aware prediction, evaluation, and decision-making.

\item \textit{Online personalization}: The nominal model is fine-tuned online using VPG-generated data during therapy, allowing the closed-loop system to adapt to evolving patient responses.

\item \textit{Real-time stochastic optimal control}: A computationally tractable sRBF-MPC framework enables nonlinear optimal control without the dense discretization required by conventional quadratic MPC.

%
\end{enumerate}

These contributions enable robust and adaptive closed-loop fluid resuscitation under clinically realistic uncertainty and variability. Unlike existing resuscitation frameworks that rely on fixed patient models or offline calibration \cite{bighamian2016lumped, estiri2023closed,estiri2024variational}, the proposed approach supports online personalization while explicitly accounting for uncertainty during both modeling and evaluation. 
This results in effective MAP regulation at a computational cost compatible with real-time implementation. To the best of the authors’ knowledge, this is the first uncertainty-aware, online personalized closed-loop fluid resuscitation framework to integrate Bayesian modeling, virtual patient generation, and real-time chance-constrained optimal control within a unified structure.

The rest of the paper is structured as follows: Section~\ref{sec2} presents the proposed uncertainty-aware modeling, control, and online adaptation framework. Section~\ref{sec3} describes the simulation setup. Section~\ref{sec4} reports open-loop model evaluation and closed-loop controller results. Section~\ref{sec5} discusses the implications, limitations, and future research directions. Finally, Section~\ref{sec6} draws the main conclusions of this work.

\section{Methodology}\label{sec2}

This section outlines the modeling and control framework developed for automated fluid resuscitation. The framework consists of two main components: (i) data-driven modeling of patient hemodynamics in response to infusion and (ii) optimal control design for personalized fluid administration. The modeling incorporates both a nominal representation of patient dynamics using the UVAE-SSM and a VPG developed through the BNSSM. The control employs an adaptive personalization strategy, using a discrete-time chance-constrained RBF optimal controller to achieve robust and individualized fluid resuscitation.

\subsection{Model Design}

The model design combines two complementary components. The UVAE-SSM captures hemodynamic responses to infusion from noisy and limited data, providing a reliable probabilistic representation of fluid resuscitation dynamics. The BNSSM develops a VPG, quantifying epistemic uncertainty, for probabilistic evaluation of closed-loop strategies. Together, these models form the basis for robust, uncertainty-aware control design.
\

The proposed modeling framework is formulated using a nonlinear state-space representation of the form
\begin{align}
& x_{k+1} = f(x_k, u_k) + v_k\label{eq1} \\
& y_k = g(x_k) + w_k
\end{align}

\noindent
where $x_k$ denotes the latent state, $u_k$ is the input, $y_k$ is the output, $v_k$ is the process noise, and $w_k$ is the measurement noise. The objective of nonlinear system identification is to infer the unknown transition and observation functions, $f(\cdot)$ and $g(\cdot)$, from measured data. These functions are approximated using the proposed UVAE-SSM and BNSSM frameworks, described next. 

\subsubsection{UVAE-SSM as Nominal Model}

The UVAE-SSM framework is developed to capture hemodynamic responses to fluid infusion. The goal is to provide a robust, control-compatible representation of patient dynamics using limited and noisy data. The UVAE-SSM models the hemodynamic system using the following nonlinear state-space representation:

\begin{equation}
\begin{aligned}
& x_{k+1} = f_{\theta}(x_k, u_k) + v_k \label{eq1} \\
& (\mu_y, \sigma_y) = g_{\phi}(x_k)\\
& y_k \sim N(\mu_y, \sigma^2_y) 
\end{aligned}
\end{equation}

\noindent
where $x_k \in \mathbb{R}^n$ represents a low-dimensional latent state that captures the underlying hemodynamic dynamics required for multi-step prediction; $u_k$ is the fluid infusion rate input and $y_k$ is the measured MAP output sampled from the predicted Gaussian distribution $\mathcal{N}(\mu_y,\sigma^2_y)$, where $\mu_y$ and $\sigma^2_y$ are the predicted MAP mean and variance. 
 The transition and observation functions,  $f_{\theta}(\cdot)$, $g_{\phi}(\cdot)$, are parameterized by neural networks with trainable parameter sets $\theta$ and $\phi$.

\begin{figure*}[!t]
  \centering
  \subfloat[]{%
      \includegraphics[width=0.48\textwidth]{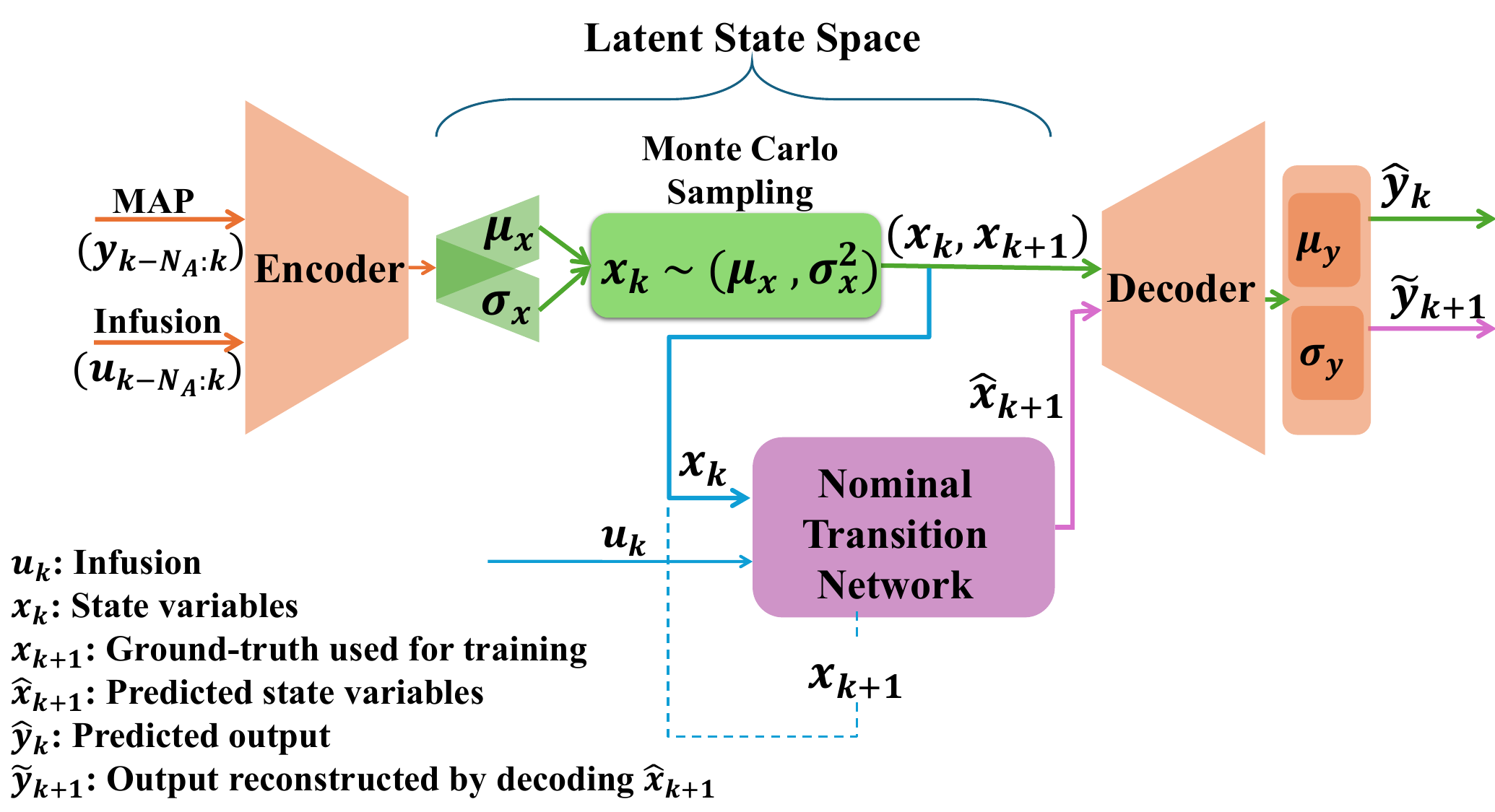}%
      \label{fig:uvae_ssm}%
  }
  \hfil
  \subfloat[]{%
      \includegraphics[width=0.48\textwidth]{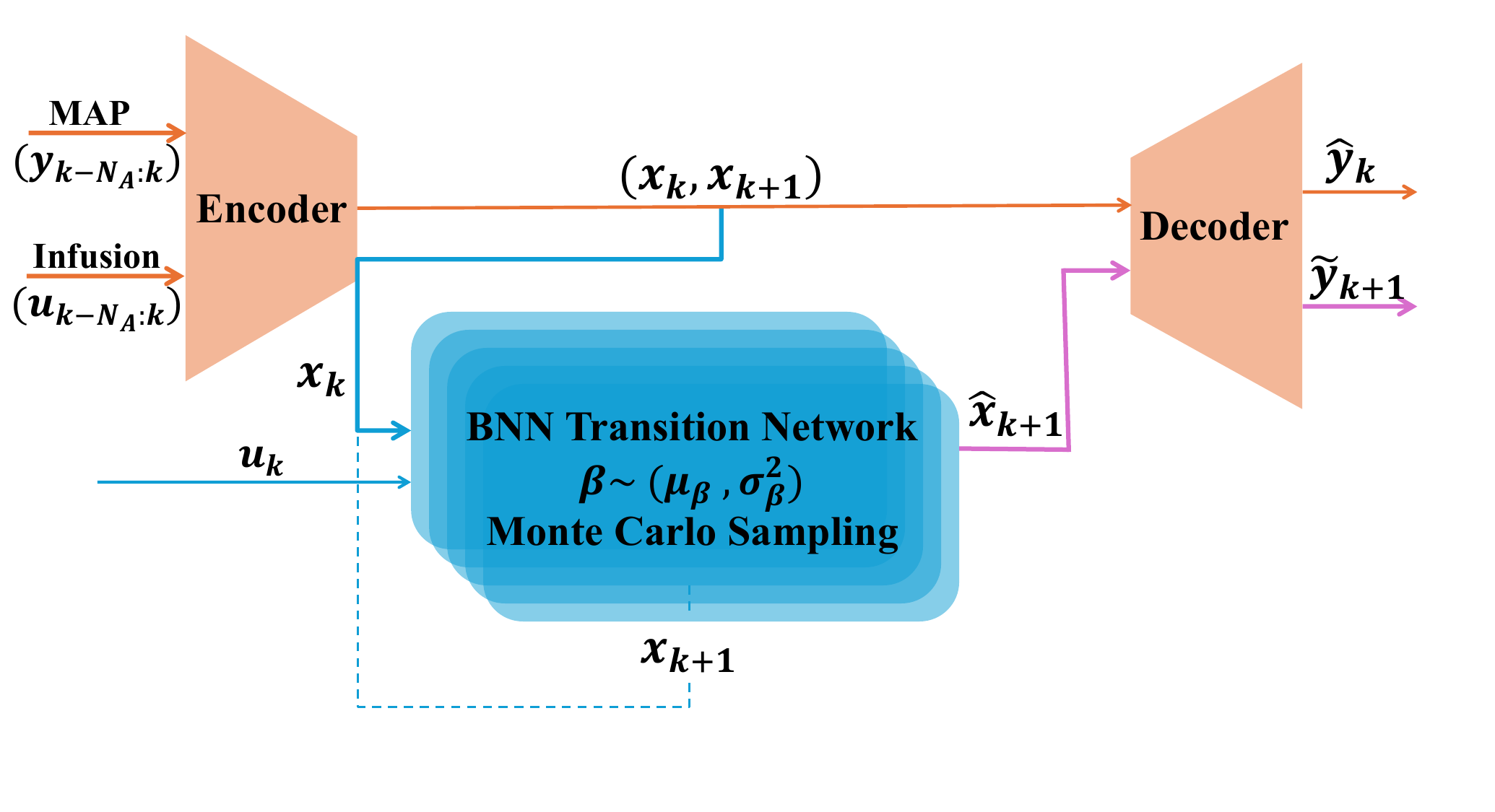}%
      \label{fig:bnssm}%
  }
  \caption{Proposed uncertainty-aware modeling framework.
  (a)~\textbf{UVAE-SSM as nominal model:} The encoder maps a sliding
  window of $N_A$ MAP measurements and infusion rates
  $(y_{k-N_A:k},\,u_{k-N_A:k})$ to the parameters $(\mu_x,\sigma_x)$
  of a Gaussian latent distribution. A latent state $x_k$ is drawn via
  the reparameterization trick and passed to the deterministic transition
  network, which predicts the next latent state $\hat{x}_{k+1}$ given
  $(x_k, u_k)$. The probabilistic decoder maps the latent state to a
  predictive MAP distribution $\mathcal{N}(\mu_y, \sigma^2_y)$, explicitly
  capturing aleatoric uncertainty in the hemodynamic output. Dashed
  lines indicate training-only connections used to provide supervision
  targets for the transition loss.
  (b)~\textbf{BNSSM as VPG}: The encoder and
  decoder form a deterministic autoencoder that maps observations to
  latent states and reconstructs MAP. The transition function
  $f_{\beta}(x_k,\,u_k)$ is implemented as a BNN, where weights $\beta$ are treated as random variables with
  variational posterior $q_{\eta}(\beta)$, approximated via Bayes by
  Backprop. At inference, $\beta$ is sampled once per complete
  simulation trajectory from $q_{\eta}(\beta)$; each distinct draw
  yields one virtual patient realization.}
  \label{fig1}
  \end{figure*}

As shown in Fig.~\ref{fig1}, the  UVAE-SSM architecture consists of three neural networks.  The encoder maps a sliding window of the $N_A$ most recent MAP measurements and infusion rates, $[y_{k-N_A:k},\, u_{k-N_A:k}]$, into the parameters of a latent state probability distribution, producing mean $\mu_x$ and standard deviation $\sigma_x$. The latent state $x_k$ is obtained by sampling from this learned distribution using the reparameterization trick, allowing aleatoric uncertainty in the physiological dynamics to be represented directly in the latent space. The transition network propagates this state forward by approximating the nonlinear function $f_{\theta}(\cdot)$, producing $x_{k+1}$. Finally, the decoder maps the latent states to the predictive MAP distribution through the observation function $g_{\phi}(\cdot)$, ensuring that the predicted outputs remain physiologically consistent with observed measurements.

\captionsetup[subfloat]{labelfont={rm,footnotesize},textfont={rm,footnotesize}}

To train the UVAE-SSM, the total loss function includes the variational autoencoder (VAE) reconstruction term and the transition consistency term, formulated to jointly learn latent representations and stable temporal dynamics. The encoder–decoder pair is first trained as a VAE to capture a compact latent representation of hemodynamics. The latent state is modeled using a multivariate Gaussian distribution, where the encoder posterior is defined as $q_\phi(x_k \mid y_{k-N_A},u_{k-N_A})=\mathcal{N}(\mu_x,\sigma_x^2)$ and the latent prior is assumed to follow a standard Gaussian distribution $p(x_k)=\mathcal{N}(0,I)$. The resulting VAE objective is given by

  \begin{equation}\label{eq3}
  \begin{aligned}
    \mathcal{L}_{\mathrm{VAE}}
    &=
    \omega_{\mathrm{nll}}\left(-\log p(y_k\mid \mu_y,\sigma_y^2)\right)\\
    &
    + \omega_{\mathrm{kl}}\,\beta_t\, D_{\mathrm{KL}}
    \!\left(
    q_\phi(x_k \mid y_{k-N_A:k},u_{k-N_A:k})
    \,\|\, \mathcal{N}(0,I)
    \right)
  \end{aligned}
  \end{equation}

\noindent

\noindent
  where $\omega_{\mathrm{nll}}$ and $\omega_{\mathrm{kl}}$ are fixed weighting coefficients that balance reconstruction quality and  regularization, $D_{\mathrm{KL}}$ denotes the Kullback--Leibler divergence between the encoder posterior and the standard Gaussian prior $\mathcal{N}(0,I)$, and $\beta_t$ is the KL annealing schedule \cite{bowman2016generating, fu2019cyclical}.

Under the Gaussian observation model, the negative log-likelihood expands as
\begin{equation}
  -\log p(y_k\mid\mu_y,\sigma_y^2) = \frac{(y_k-\mu_y)^2}{2\sigma_y^2}
  + \frac{1}{2}\log(2\pi\sigma_y^2)
  \end{equation}
where the first term penalizes the weighted squared deviation of the predicted mean from the observed MAP, and the second term regularizes the predicted variance, preventing the model from inflating uncertainty to minimize reconstruction error \cite{kendall2017uncertainties}.  The Gaussian prior is selected to promote a smooth and structured latent space while enabling stable transition learning and efficient reparameterization-based optimization. The probabilistic decoder explicitly captures aleatoric uncertainty in MAP prediction.

Using the latent trajectories extracted from the trained VAE, the transition network is then optimized to capture the temporal evolution of latent states, with internal consistency, enforced through the encoder and decoder. The transition loss is given by
\begin{equation}\label{eq4}
\mathcal{L}_{\mathrm{transition}} = \omega_t L_x(\hat{x}_{k+1}, x_{k+1}) + \omega_o L_y(\tilde{y}_{k+1}, y_{k+1})
\end{equation}

\noindent 
where $x_{k+1}$ denotes the target latent state obtained by encoding the observed measurements at time step $k+1$ and serves as the ground-truth target for training the transition network, $\hat{x}_{k+1}$ is the latent state predicted by the transition network, and $\tilde{y}_{k+1}$ is the MAP reconstructed by decoding $\hat{x}_{k+1}$. The coefficients $\omega_t$ and $\omega_o$ are weighting factors that determine the relative contributions of the latent-state prediction loss and the output reconstruction loss, respectively. The first term enforces accurate latent prediction and the second constrains decoder consistency with the observed MAP. 


By jointly minimizing $\mathcal{L}_{\mathrm{VAE}}$ and $\mathcal{L}_{\mathrm{transition}}$, the UVAE-SSM maintains agreement between latent predictions and decoded MAP probability, thereby enhancing robustness against error propagation and ensuring reliable performance in real-time control applications.
\\

\subsubsection{BNSSM as VPG}

The BNSSM is developed to quantify epistemic uncertainty, which arises from limited data and incomplete knowledge of patient-specific physiological dynamics. The resulting model serves as the foundation for the proposed VPG, enabling probabilistic evaluation of closed-loop control strategies across inter- and intra-patient variability prior to clinical deployment. The BNSSM adopts the following nonlinear state-space representation:

\begin{equation}
\begin{aligned}
& x_{k+1} = f_{\beta}(x_k, u_k) \\
& y_k = g_{\psi}(x_k)
\end{aligned}
\end{equation}

\noindent
\noindent
where $g_{\psi}(\cdot)$ represents a deterministic decoder parameterized by $\psi$, and $f_{\beta}(\cdot)$ denotes the BNN transition function, whose weight parameters $\beta$ are treated as random variables representing epistemic uncertainty.  The posterior distribution over the BNN weights is approximated using a variational distribution
\begin{equation}
q_{\eta}(\beta)
=
\mathcal{N}(\mu_{\beta},
\sigma_{\beta}^{2})
\end{equation}
where $\mu_{\beta}$ and $\sigma_{\beta}^{2}$ denote the mean and variance of the posterior distribution, respectively, and $\eta=\{\mu_{\beta},\sigma_{\beta}\}$ represents the set of variational parameters.

The predictive distribution of the latent state at time step $k+1$ is expressed as
\begin{equation}
\begin{aligned}
p(x_{k+1}\mid x_k,u_k,D)
&= \int p(x_{k+1}\mid x_k,u_k,\beta)q_{\eta}(\beta)\, d\beta
\end{aligned}
\end{equation}

\noindent
where the integral marginalizes predictions over all possible weight configurations.
In practice, this expectation is estimated through Monte Carlo sampling, where multiple draws of $\beta$ from $q_{\eta}(\beta)$ generate an ensemble of trajectories, and each sample produces one distinct virtual patient realization. This structure enables the model to capture variability in physiological responses arising from epistemic uncertainty.

  The BNSSM is trained in two sequential stages:  First, the encoder
  and decoder are jointly optimized to minimize the autoencoder (AE)
  reconstruction loss
  \begin{equation}
    \mathcal{L}_{\mathrm{AE}}
    = L_{y}(\hat{y}_{k},\,y_{k})
  \end{equation}
  where $L_{y}(\hat{y}_{k},y_{k})$ is the mean squared error between the
  measured MAP, $y_k$, and its reconstruction, $\hat{y}_{k}=g_{\psi}(x_k)$.
 Second, the BNN transition network is trained by minimizing the variational objective
  \begin{equation}\label{bnnloss}
  \begin{aligned}
    \mathcal{L}_{\mathrm{BNN\text{-}trans}}
    &= \mathbb{E}_{q_{\eta}(\beta)}
         \!\bigl[\mathcal{L}_{\mathrm{transition}}\bigr]
     + \lambda_{\mathrm{KL}}\,
       D_{\mathrm{KL}}\!\bigl(
         q_{\eta}(\beta)\,\|\,p(\beta)
       \bigr)
  \end{aligned}
  \end{equation}
  where $\mathcal{L}_{\mathrm{transition}}$ is defined in~\eqref{eq4} and
  $\lambda_{\mathrm{KL}}$ is a positive weighting factor controlling the
  strength of Bayesian regularization.  This objective follows the
  Bayes-by-Backprop framework~\cite{blundell2015weight}. The first term is
  the expected transition prediction loss and the second term is the KL
  complexity cost that regularizes $q_{\eta}(\beta)$ toward the prior.

The KL term regularizes the variational posterior toward the prior to ensure calibrated uncertainty in the transition weights. It is defined as
\begin{equation}
D_{\mathrm{KL}}\big(q_{\eta}(\beta)|p(\beta)\big)
= \mathbb{E}_{q_{\eta}(\beta)}\big[\log q_{\eta}(\beta) - \log p(\beta)\big]
\end{equation}

\noindent where $p(\beta)$ denotes the prior distribution over the transition network weights. During training, the expectation over $q_{\eta}(\beta)$ is approximated with $T$ Monte Carlo samples using the reparameterization

\begin{equation}
\beta^{(t)} = \mu_{\beta} + \sigma_{\beta} \odot \varepsilon^{(t)}, \qquad
\varepsilon^{(t)} \sim \mathcal{N}(0, I)
\end{equation}

\noindent where $t = 1, \ldots, T$ indexes the Monte Carlo samples drawn from the variational posterior.
This formulation maintains differentiability of the stochastic objective, enabling efficient gradient estimation through backpropagation.
As a result, the network learns a distribution of transition functions rather than a single deterministic mapping, providing a principled mechanism for epistemic uncertainty quantification in patient modeling.

\subsection{Controller Design}

  This section presents the discrete-time stochastic RBF optimal control strategy for automated fluid resuscitation. Building upon the UVAE-SSM and BNSSM models, the controller is designed to regulate
  MAP within physiological limits while accounting for inter- and intra-patient variability. To account for predictive uncertainty, the UVAE-SSM probabilistic output is embedded directly into the control constraints through a chance-constraint formulation, yielding a safety-aware closed-loop policy.

\subsubsection{Discrete-Time Stochastic RBF Optimal Controller}

  We formulate automated fluid resuscitation as a finite-horizon discrete-time
  optimal control problem by embedding the learned UVAE-SSM dynamics into the control design.
  The objective is to regulate MAP to a desired reference while minimizing
  control effort over the prediction horizon. The resulting optimization
  problem is defined as
  \begin{equation}
  \label{eq:ocp}
  J = \sum_{k=0}^{N_h-1} L(y_k, u_k)
  \end{equation}
  subject to
  \begin{align}
  & x_{k+1} = f_{\theta}(x_k, u_k)+v_k \label{eq:dyn}\\
  & (\mu_y, \sigma_y) = g_{\phi}(x_k)\label{eq:dis}\\
  & y_k \sim \mathcal{N}(\mu_y,\,\sigma^2_y)
    \label{eq:obs}\\
  & \lambda(x[0], x_0, x[N], x_f) = 0 \label{eq:bc}\\
  & \mu_y - z_{1-\alpha}\,\sigma_y \geq y_{\min}
    \label{eq:cc}
  \end{align}

\noindent
where $L(\cdot)$ is a running cost penalizing MAP tracking error and control effort at each time step, $y_{\min}$ is the physiological safety floor, boundary constraints $\lambda(\cdot)$ enforce consistency with the initial condition, and constraint~\eqref{eq:cc} encodes the probabilistic physiological safety requirement, as derived below. 

  Since the UVAE-SSM probabilistic decoder provides a Gaussian distribution
  over the predicted MAP at each step, the physiological safety floor is
  enforced via a chance constraint. The requirement that MAP remains above
  $y_{\min}$ with probability of at least $1-\alpha$ is stated as
  \begin{equation}
    P\!\left(y_k \geq y_{\min}\right) \geq 1 - \alpha,
    \quad k = 0,\ldots,N_h-1.
    \label{eq:chance}
  \end{equation}
  Under the Gaussian model in~\eqref{eq:obs},
  this admits the closed-form deterministic reformulation in
  constraint~\eqref{eq:cc}, where
  $z_{1-\alpha}=\Phi^{-1}(1-\alpha)$ is the $(1-\alpha)$-quantile of the
  standard normal distribution, and $\Phi^{-1}(\cdot)$ denotes the cumulative distribution function of the standard
  normal distribution $\mathcal{N}(0,I)$. When predictive uncertainty $\sigma_y$ is large, the effective safety
  margin expands automatically, making the controller inherently conservative
  in proportion to its prediction confidence.

  The RBF framework leverages global basis functions to approximate nonlinear
  state and control trajectories parameterized as
  \begin{align}
  x_k &\approx x^R_k = \sum_{i=1}^{N_b} \alpha_i\,\varphi_i[k]\\
  u_k &\approx u^R_k = \sum_{i=1}^{N_b} \gamma_i\,\varphi_i[k]
  \end{align}
  where $\alpha_i$ and $\gamma_i$ are optimization parameters, and
  $\varphi_i[k]$ is defined as a Gaussian RBF,
  \begin{equation}
  \varphi_i[k] = \exp\!\left(-\epsilon^2\,\|k-c_i\|^2\right)
  \end{equation}
  with center $c_i$ and fixed shape parameter
  $\epsilon$, a design hyperparameter that controls the spread
  of the kernel. The index $k$ denotes the discrete time point. 
Gaussian RBFs were selected because they provide smooth control trajectories and high parameterization accuracy, and they have been used in optimal control applications such as medication dosing \cite{mirinejad2017individualized, mirinejad2021radial, mirinejad2015rbf,  mirinejad2015radial}.

  Forward-shifting the RBF basis functions with respect to the
  discrete index yields the state trajectory:
  \begin{equation}
  x^R_{k+1} \approx \sum_{i=1}^{N_b} \alpha_i\,\varphi_i[k+1].
  \end{equation}

    This parameterization reformulates the state-space model
    in~\eqref{eq:dyn} and~\eqref{eq:dis} into the following RBF-based
    representation:
    \begin{align}
    &x^{R}_{k+1} = f\!\left(\alpha_i,\gamma_i,\varphi_i[k_j]\right)\\
    & (\mu_y^R, \sigma_y^R) = g_{\phi}(x_k^R)
    \end{align}

    Substituting these expressions into the dynamic constraint~\eqref{eq:dyn}
    produces an algebraic collocation condition:
    \begin{equation}
    \sum_{i=1}^{N_b}\alpha_i\,\varphi_i[k_j+1]
    - f\!\left(\alpha_i,\gamma_i,\varphi_i[k_j]\right) = 0,
    \quad j=1,\dots,N_b.
    \end{equation}

    The fluid resuscitation problem is then posed as a finite-horizon optimal
    control problem:
    \begin{equation}\label{eq16}
    \begin{aligned}
    \min_{A,B}\;J = \sum_{k=0}^{N_h-1}\Big(
      &Q\,\|\mu_y^R-y_{\mathrm{ref}}\|^2
      + R\,\|u^R_k\|^2\\
      &+ S\,\|\Delta u^R_k\|^2
    \Big)
    \end{aligned}
    \end{equation}
    subject to the dynamics~\eqref{eq:dyn},
    boundary conditions~\eqref{eq:bc},
    input bounds $0\leq u^R_k\leq u_{\max}$,
    and the chance constraint~\eqref{eq:cc}. $Q>0$, $R>0$, and $S>0$ are weights for the decoder-predicted
    mean MAP tracking error, fluid infusion rate, and infusion rate
    of change, respectively; $y_{\mathrm{ref}}$ is the MAP traget, and $N_h$ is the prediction horizon. The resulting discrete-time optimal control problem is solved as a nonlinear programming (NLP) problem to find $A=(\alpha_1,\ldots,\alpha_{N_b})^T$ and $B=(\gamma_1,\ldots,\gamma_{N_b})^T$, the state and control expansion parameters, using the sequential least squares programming (SLSQP) solver available in \texttt{scipy.optimize.minimize}. 

\subsubsection{Online Adaptive Personalization for Fluid Resuscitation}

  While UVAE-SSM provides a reliable,
  nominal model, it does not account for patient-specific dynamics and is therefore insufficient for guiding personalized therapy. To address this limitation, an online adaptive framework is implemented, where the nominal model is progressively fine-tuned using data generated during treatment. This process enables continuous personalization as therapy progresses, allowing the controller to evolve from a generic to a patient-specific
  representation of hemodynamics. Throughout this process, the
  controller does not have access to the true patient dynamics at any
  point. In other words, sRBF-MPC operates exclusively on the UVAE-SSM as its internal predictive  model, while BNSSM serves as the simulated patient environment (VPG). The adaptation proceeds in three sequential
  phases:

  \begin{itemize}
      \item \textbf{Phase 1 (0--30~min):} The subject undergoes
      hemorrhage during the first 15~minutes without treatment, followed by
      a period of no infusion representing the pre-hospital stage. 
      The nominal UVAE-SSM remains fixed and serves
      as the nominal model for the controller.

      \item \textbf{Phase 2 (from 30~min):} Closed-loop
      fluid resuscitation is initiated at $t = 30$~min, where the
      sRBF-MPC operates using the nominal UVAE-SSM within a receding horizon MPC framework.

      \item \textbf{Phase 3 (periodic, starting at t=60~min):} Continuous adaptation is maintained by periodically fine-tuning the UVAE-SSM transition network using the most recent closed-loop data from VPG. The first fine-tuning step is performed at minute 60, followed by updates at 30-minute intervals (i.e., 60, 90, 120, and 150 minutes). This rolling update progressively personalizes the predictive model.
  \end{itemize}

  This online adaptive framework enables the controller to transition from a nominal to an individualized predictive model, ensuring robust and patient-specific control.

  \begin{figure}[!t]
  \centering
  \includegraphics[width=0.5\textwidth]{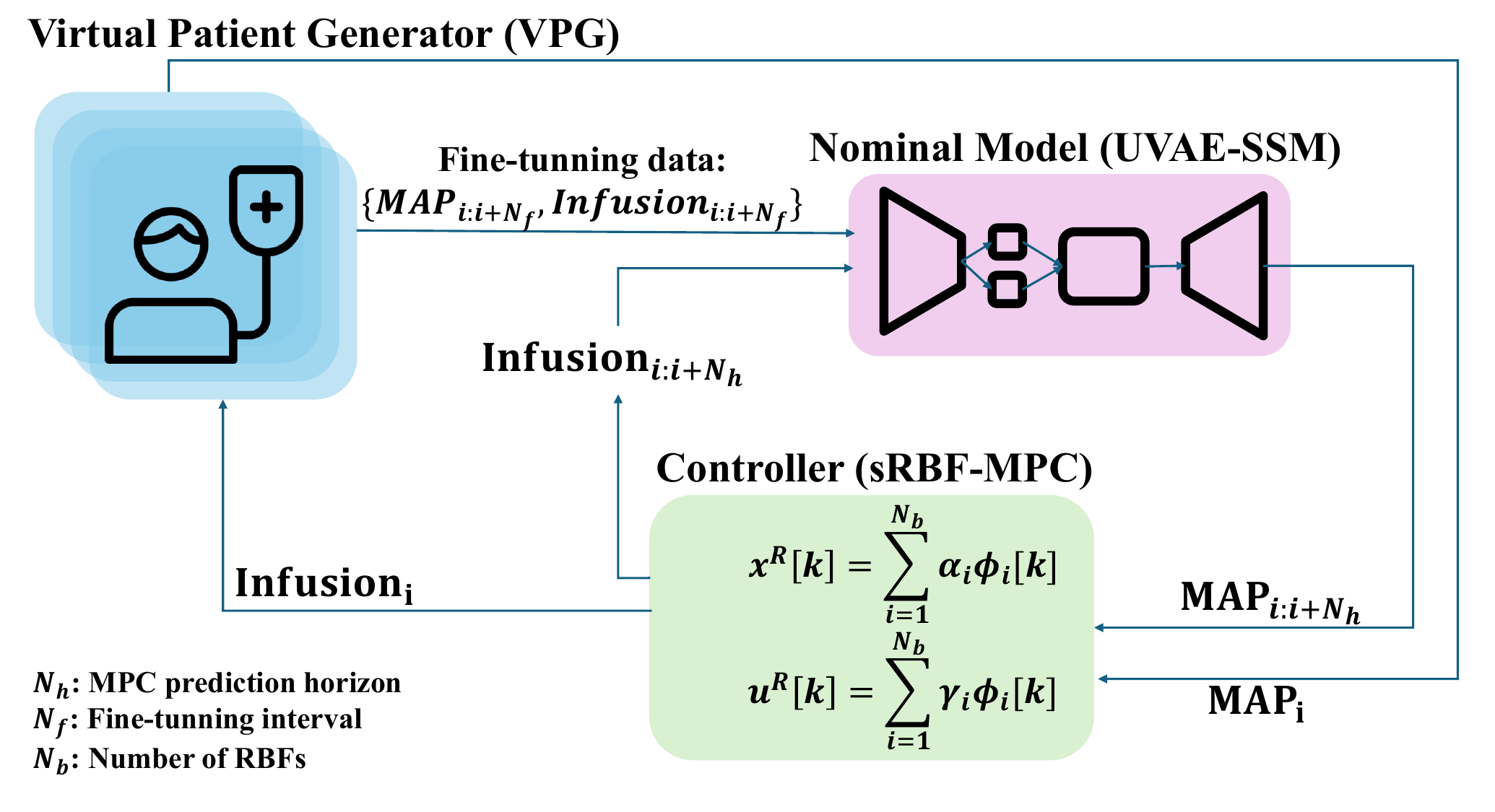}
  \caption{Structure of the closed-loop fluid resuscitation framework with online personalization.}
  \label{fig3}
  \end{figure}

The structure of the closed-loop fluid resuscitation framework with online personalization is
  illustrated in Fig.~\ref{fig3}. The sRBF-MPC operates within a receding-horizon MPC framework. At each sampling instant, the current MAP measurement is provided to the controller, and the UVAE-SSM predicts the future MAP trajectory over the prediction horizon. The sRBF-MPC then computes the optimal fluid infusion rate by solving a finite-horizon optimization problem subject to the physiological safety constraints and the chance constraint~\eqref{eq:cc}. Only the first element of the optimized infusion sequence is applied to the simulated patient (VPG), and the resulting MAP response is fed back to the controller. This closed-loop interaction repeats at every time step, while the UVAE-SSM transition network is periodically fine-tuned through the online personalization mechanism described above.

  \section{Simulation Setup}\label{sec3}

  This section describes the dataset, model training procedures, controller
  implementation, personalization strategy, and evaluation methodology used
  to assess the proposed framework. The UVAE-SSM and BNSSM were trained
  on animal data and evaluated on both held-out animal subjects
  and an independent human dataset. Closed-loop controller evaluation was
  performed in simulation using the VPG as the patient environment. All
  models were implemented in PyTorch.


  The hemorrhage resuscitation dataset used in this work was obtained from a
  sheep  animal study approved by the
  Institutional Animal Care and Use Committee at the University of Texas
  Medical Branch~\cite{rafie2004hypotensive}.
  Data from 16 sheep, comprising time-series measurements of
  MAP, fluid infusion rates, and hemorrhage profiles, were used in this work.
  Each animal subject underwent a controlled blood loss protocol, consisting of an
  initial hemorrhage of 25~ml/kg during the first 15~minutes, followed by
  additional losses of 5~ml/kg at 50 and 70~minutes.
  Fluid resuscitation with lactated Ringer's solution, a
  crystalloid fluid,  began at 30~minutes and continued for the
  remainder of the 180-minute experiment.
  Fourteen subjects were used to train the UVAE-SSM and BNSSM and two subjects were held out for evaluation of both models and were not used in any training stage.

  To assess cross-population generalizability, the UVAE-SSM and BNSSM were
  further evaluated on an independent human clinical data cohort\cite{drobin1999volume}.
 In that study, 10 healthy human subjects underwent three intravenous infusion experiments with 25 mL/kg of lactated Ringer's solution administered over 30 minutes: once while normovolemic, once after withdrawal of 450 mL of venous blood (mild hypovolemia), and once after withdrawal of 900 mL (moderate hypovolemia), yielding 30 experimental sessions across the three conditions. Hemodynamic variables, including systolic and diastolic blood pressure (SBP and DBP), were recorded at regular intervals, with blood sampling continuing for up to 180 minutes after the start of infusion, from which MAP was derived using the standard approximation MAP = 1/3 SBP + 2/3 DBP. Six sessions spanning all three hypovolemia conditions were selected for evaluation. No human data were used for model training.



The UVAE-SSM consisted of an encoder, a latent transition model, and a probabilistic decoder. The encoder used a temporal window of $N=20$ measurements and a latent dimension of $n_x=6$, while the decoder output both MAP mean and variance to quantify aleatoric uncertainty. Training followed a three-stage curriculum consisting of VAE pretraining, transition-consistency learning, and long-horizon rollout training. 

During the VAE pretraining phase, the encoder and decoder were trained jointly to learn a compact and consistent latent representation of the observed dynamics. In~\eqref{eq3}, loss weights were set to $\omega_{\mathrm{nll}}=10.0$ and $\omega_{\mathrm{kl}}=1.0$, and the KL coefficient $\beta_t$ was annealed from 0 to 1 over the first 120 epochs to balance reconstruction and regularization. 

In the transition-consistency phase, the latent transition model was trained while maintaining consistency with the encoder representation. 
The loss weights in~\eqref{eq4} were selected as $\omega_t=1$ and $\omega_o=0.3$ to encourage accurate latent-state propagation while maintaining consistency with the encoder. In the final rollout training phase, the latent transition model was optimized to recursively use its own predictions over a 5-minute horizon, enabling it to learn stable and consistent multi-step forecasting behavior.

Optimization was performed using Adam with weight decay, gradient clipping, and adaptive learning-rate scheduling. Overfitting was mitigated through KL regularization, probabilistic reconstruction loss, and the staged training procedure. Predictive uncertainty was estimated using $T=50$ Monte Carlo samples and reported as $\pm2\sigma$ confidence bands. Model performance was evaluated on two held-out sheep subjects, compared against a lumped hemodynamic model~\cite{bighamian2018control}, and further assessed on six held-out human subjects to evaluate cross-population generalizability.


The BNSSM employed the same encoder and decoder architecture as the UVAE-SSM while introducing epistemic uncertainty through a Bayesian transition network with Gaussian weight distributions. Training followed the same three-stage curriculum, with Bayesian regularization enforced through the KL-divergence term in~(\ref{bnnloss}). During inference, epistemic uncertainty was quantified using $T=50$ Monte Carlo weight realizations, each generating a distinct virtual patient trajectory.


The sRBF-MPC controller was implemented using the UVAE-SSM as the internal predictive model. The prediction horizon was set to $N_h = 20$ steps, and the control trajectory was parameterized using $N_b = 10$ Gaussian RBFs with $\epsilon= 0.5$. The basis functions were centered at uniformly spaced locations across the prediction horizon.

The controller objective was to regulate MAP to a target value of 80 mmHg while minimizing infusion effort and changes in infusion rate. The corresponding cost weights were set to $Q = 3.0$, $R = 0.1$, and $S = 0.01$. Safety was enforced through chance constraints requiring the predicted MAP to remain above 60 mmHg with a probability of at least 95\% throughout the prediction horizon. Predicted uncertainty was propagated using Monte Carlo sampling, and the resulting constraint satisfaction probability was incorporated into the optimization problem. The infusion rate was constrained to $u \in [0,100]$ ml/min \cite{mirinejad2019evaluation}. At each control step, the resulting nonlinear optimization problem was solved using SLSQP (\texttt{scipy.optimize.minimize}). 

Closed-loop fluid resuscitation was initiated at 30 minutes. To account for evolving patient dynamics, the UVAE-SSM transition network was subsequently fine-tuned every 30 minutes using the most recent 30 minutes of MAP and infusion data generated by the VPG after hospitalization. At each adaptation event, the transition  network was updated for 12 epochs using Adam with a learning rate of $2\times10^{-4}$. Only the transition network was adapted online, while the encoder and decoder remained fixed. 


The sQ-MPC and Q-MPC were implemented as baseline controllers using the same UVAE-SSM predictive model, objective function, input bounds, and SLSQP optimization framework as the proposed sRBF-MPC. The sQ-MPC employed the same chance-constraint formulation but directly optimized the full $N_h=20$ control sequence, whereas the Q-MPC optimized the same control sequence without chance constraints. These comparisons isolate the effects of the RBF parameterization and chance-constrained formulation on closed-loop performance.


Model and control performance were evaluated using root mean squared error (RMSE), mean absolute error (MAE), and median absolute percentage error (MDAPE). Closed-loop performance was further assessed using total infusion volume and B60, defined as the percentage of time MAP remained below the 60 mmHg safety threshold.

\section{Results}\label{sec4}

\begin{figure*}[!t]                                      \centering                                            \includegraphics[width=\textwidth]{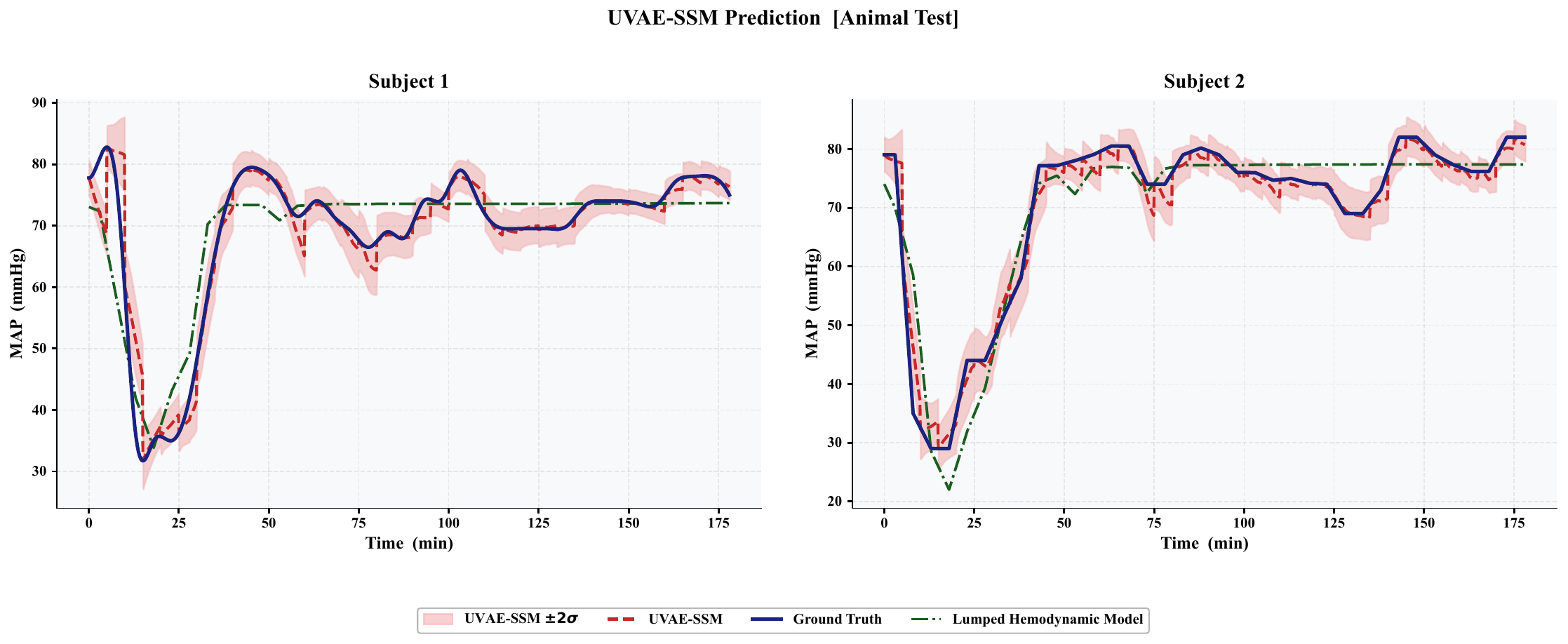}                  \caption{MAP prediction for held-out animal test subjects using the proposed UVAE-SSM and the lumped hemodynamic model \cite{bighamian2018control}.}
\label{fig:uvae_animal}
\end{figure*}

\begin{figure*}[!t]  
      \centering                                                         \includegraphics[width=\textwidth]{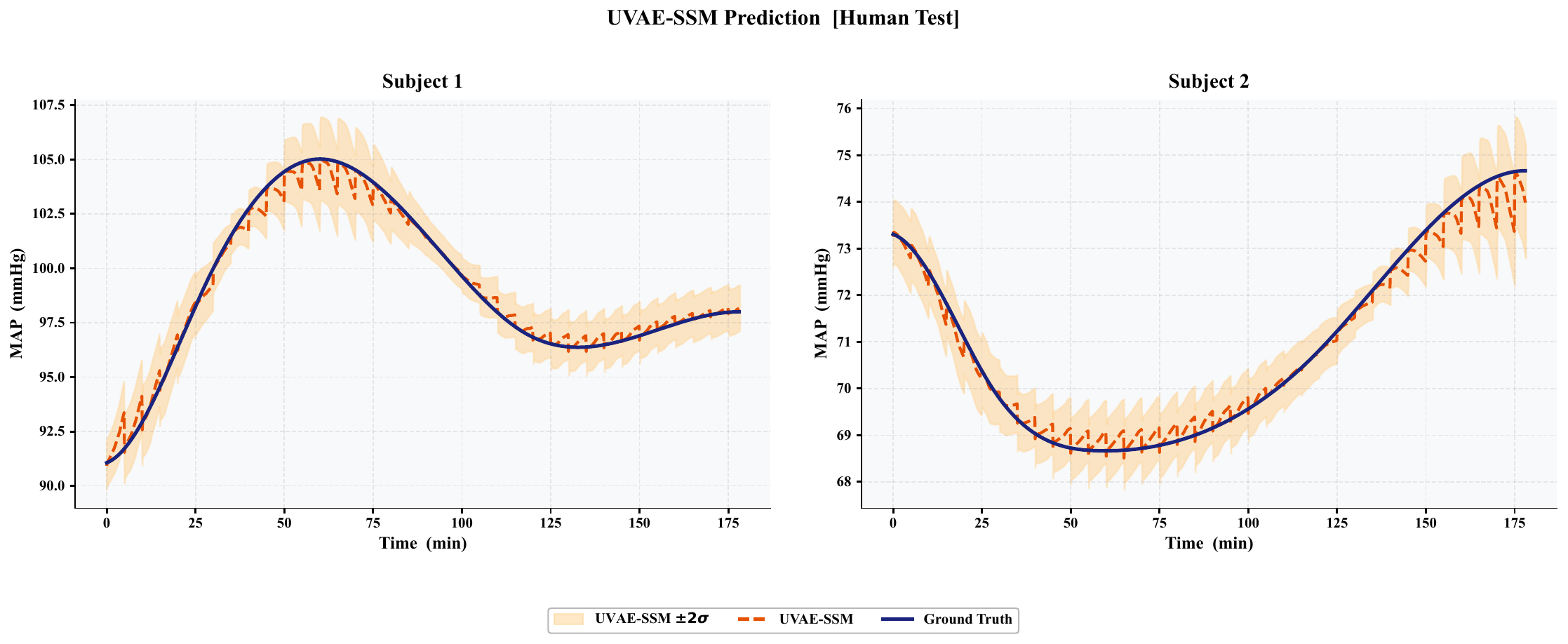}
      \caption{MAP prediction for human subjects using UVAE-SSM.}
      \label{fig:uvae_human}
\end{figure*}

\textit{Model Evaluation:} Figs ~\ref{fig:uvae_animal} and~\ref{fig:uvae_human} show MAP predictions generated by the UVAE-SSM for the held-out animal and human cohorts, respectively. For the animal subjects, the UVAE-SSM accurately reproduced the major MAP variations induced by hemorrhage and fluid resuscitation while maintaining tight uncertainty bounds throughout the 180-minute rollout. In contrast, the lumped hemodynamic model exhibited larger deviations from the measured trajectories. For the human subjects, the UVAE-SSM closely tracked the measured MAP trajectories despite being trained exclusively on animal data, demonstrating strong cross-population generalizability.

Quantitative results are summarized in Table~\ref{tab:prediction_metrics_UVAE_SSM}. For the animal subjects, the UVAE-SSM achieved RMSE values of 4.66\% and 3.07\%, substantially outperforming the lumped hemodynamic model, which produced RMSE values of 7.74\% and 7.19\% on the same subjects. On the human cohort, RMSE values ranged from 0.53\% to 2.71\%, with MAE values below 1.8\% and MDAPE values below 1.2\%. These results demonstrate that the learned latent representation captures physiologically meaningful dynamics that remain informative across both animal and human datasets.

\begin{figure*}[!t]
\centering
\includegraphics[width=\textwidth]{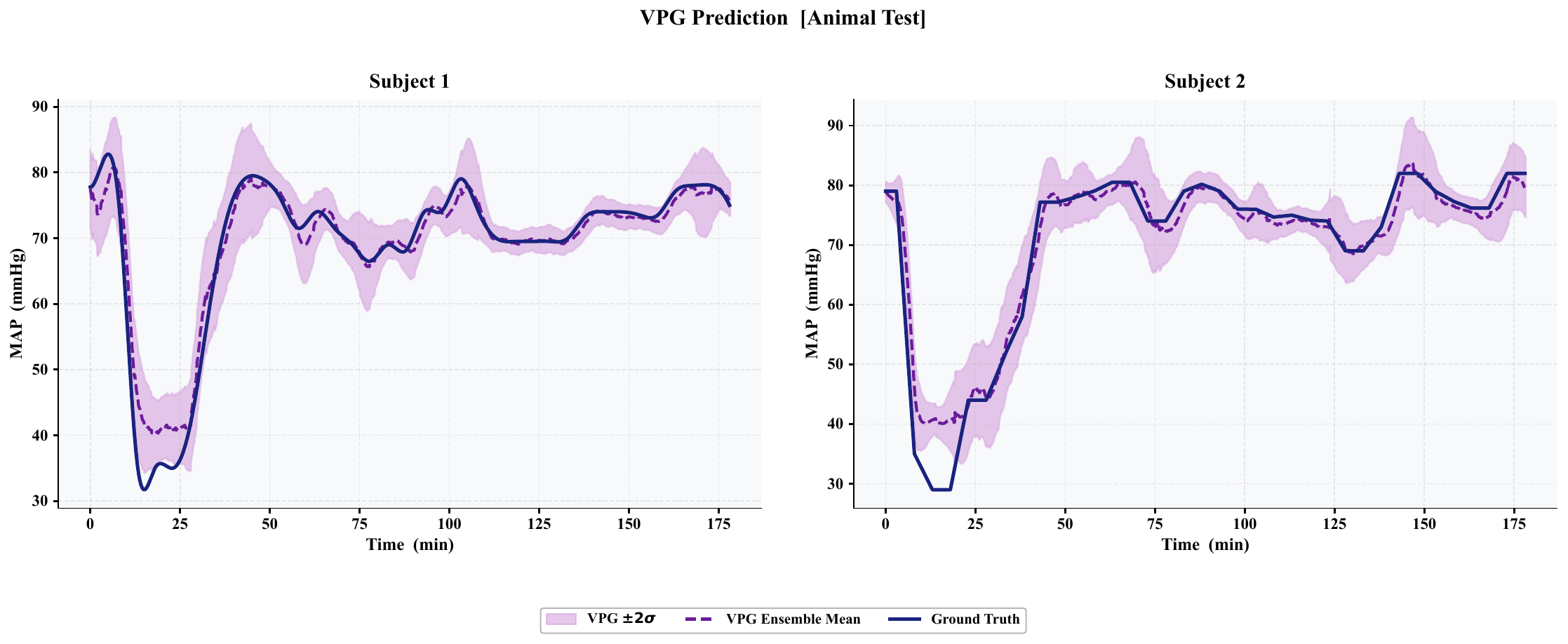}      
\caption{MAP prediction on held-out animal test subjects using the VPG. The dashed line shows the ensemble mean over $N=50$ virtual patients.} 
\label{fig:vpg_animal} 
\end{figure*}

\begin{figure*}[!t]
\centering
\includegraphics[width=\textwidth]{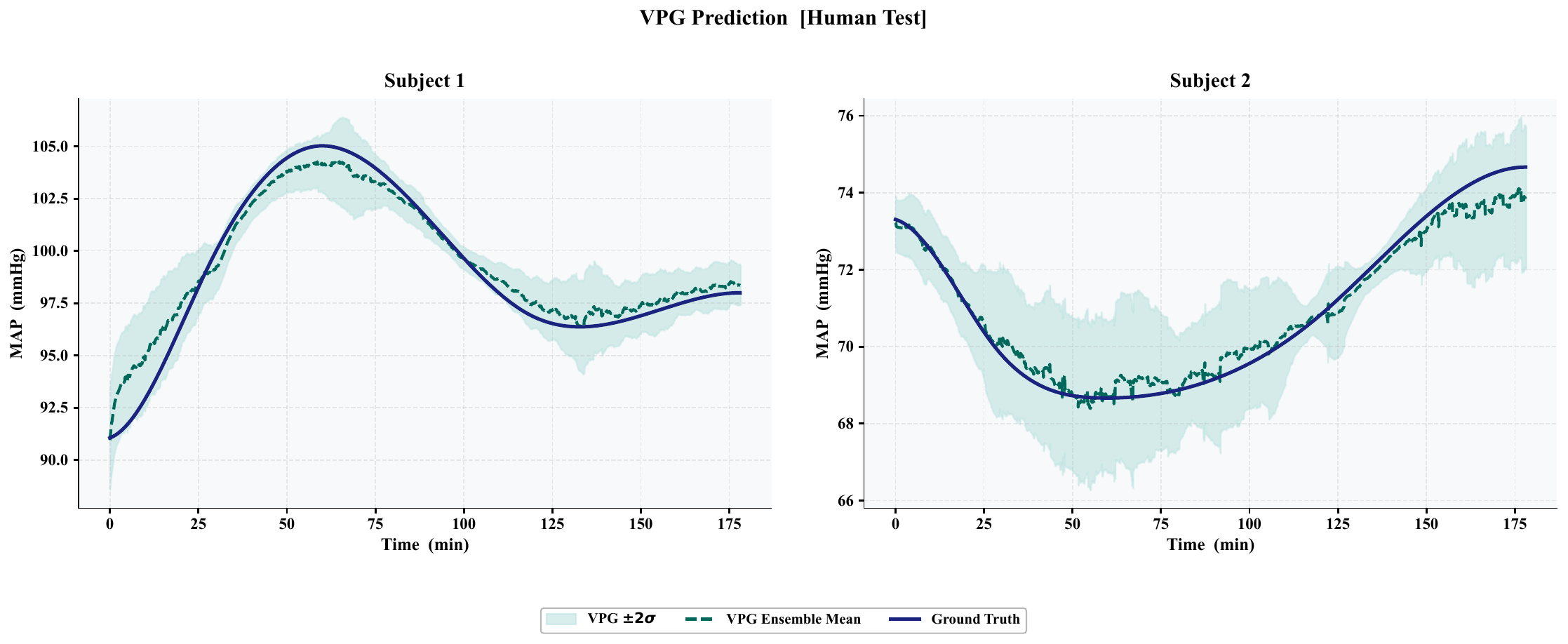}      
\caption{MAP prediction on held-out human clinical test subjects using the VPG.} 
\label{fig:vpg_human} 
\end{figure*}

Figs~\ref{fig:vpg_animal} and~\ref{fig:vpg_human} show MAP predictions generated by the VPG for the same held-out animal and human subjects. In both cohorts, the VPG accurately reproduced the observed MAP dynamics while providing epistemic uncertainty estimates through an ensemble of $50$ virtual patients. The uncertainty bands capture variability across the ensemble trajectories and reflect uncertainty in the learned transition dynamics.

Quantitative results are summarized in Table~\ref{tab:vpg_metrics}. For the held-out animal subjects, the VPG achieved RMSE values of 6.91\% and 3.35\%, with corresponding MDAPE values below 2\%. On the human cohort, RMSE values ranged from 0.42\% to 2.38\%, with similarly low MAE and MDAPE values. The low prediction errors observed across both animal and human subjects indicate that the BNSSM preserves predictive accuracy while explicitly quantifying epistemic uncertainty, enabling the generation of realistic virtual patient populations for robust controller evaluation.

\begin{table*}[!h]
    \centering
    \caption{MAP prediction performance of the UVAE-SSM on held-out animal and human test subjects. The lumped hemodynamic model~\cite{bighamian2018control} is included as a physiologically grounded baseline for the animal subject evaluation.}
    \label{tab:prediction_metrics_UVAE_SSM}
    \begin{tabular}{l cc cc cccccc}
    \toprule
    & \multicolumn{2}{c}{\textbf{UVAE-SSM (Animal)}}
    & \multicolumn{2}{c}{\textbf{Lumped Model (Animal)}}
    & \multicolumn{6}{c}{\textbf{UVAE-SSM (Human)}} \\
    \cmidrule(lr){2-3} \cmidrule(lr){4-5} \cmidrule(lr){6-11}
    \textbf{Metric}
      & \textbf{Subject1} & \textbf{Subject2}
      & \textbf{Subject1} & \textbf{Subject2}
      & \textbf{Subject1} & \textbf{Subject2} & \textbf{Subject3}
      & \textbf{Subject4} & \textbf{Subject5} & \textbf{Subject6} \\
    \midrule
    RMSE (\%)
      & 4.66 &  3.07
      & 7.74 & 7.19
      & 2.49 & 0.86 & 0.91 & 2.71 & 0.54 & 0.53 \\
    MAE (\%)
      & 2.10 &  1.88
      & 5.84 & 5.38
      & 1.76 & 0.59 & 0.44 & 1.14 & 0.38 & 0.38 \\
    MDAPE (\%)
      & 0.65 &  1.02
      &  5.57 &  3.81
      & 1.13 & 0.39 & 0.23 & 0.49 & 0.25 & 0.26 \\
    \bottomrule
    \end{tabular}
  \end{table*}

  \begin{table*}[!t]
    \centering
    \caption{MAP prediction performance of the VPG on held-out animal and human test subjects}
    \label{tab:vpg_metrics}
    \begin{tabular}{l cc cccccc}
    \toprule
    & \multicolumn{2}{c}{\textbf{VPG (Animal)}}
    & \multicolumn{6}{c}{\textbf{VPG (Human)}} \\
    \cmidrule(lr){2-3} \cmidrule(lr){4-9}
    \textbf{Metric}
      & \textbf{Subject1} & \textbf{Subject2}
      & \textbf{Subject1} & \textbf{Subject2} & \textbf{Subject3}
      & \textbf{Subject4} & \textbf{Subject5} & \textbf{Subject6} \\
    \midrule
    RMSE (\%)
      & 6.91 & 3.35
      & 2.38 & 0.87 & 0.54 & 0.63 & 0.42 & 0.57 \\
    MAE (\%)
      & 4.01 & 2.04
      & 2.04 & 0.60 & 0.43 & 0.42 & 0.31 & 0.42 \\
    MDAPE (\%)
      & 1.84 & 1.13
      & 1.71 & 0.32 & 0.40 & 0.20 & 0.21 & 0.33 \\
    \bottomrule
    \end{tabular}
  \end{table*}

\textit{Controller Evaluation}: Four hemorrhage scenarios were designed to evaluate controller robustness across a
  range of physiological conditions. Each scenario used a unique virtual patient from the VPG, producing different patient-specific dynamics while keeping the underlying
  model fixed. Scenarios~1--3 each involved a single primary hemorrhage event spanning
  $t = 0$--$15$~min, with hemorrhage rates of 4, 2, and 3~mmHg/min, respectively.
  Scenario~4 involved a primary hemorrhage at 4.5~mmHg/min ($t = 0$--$15$~min), followed
  by a secondary hemorrhage at 4~mmHg/min ($t = 80$--$87$~min) occurring during the active control phase. In all scenarios, controller activation began at $t = 30$~min. Hemorrhage was introduced as an
  external disturbance through perturbations of the VPG latent state designed to induce
  controlled reductions in MAP.

\begin{figure*}[!t]
\centering
\includegraphics[width=\textwidth]{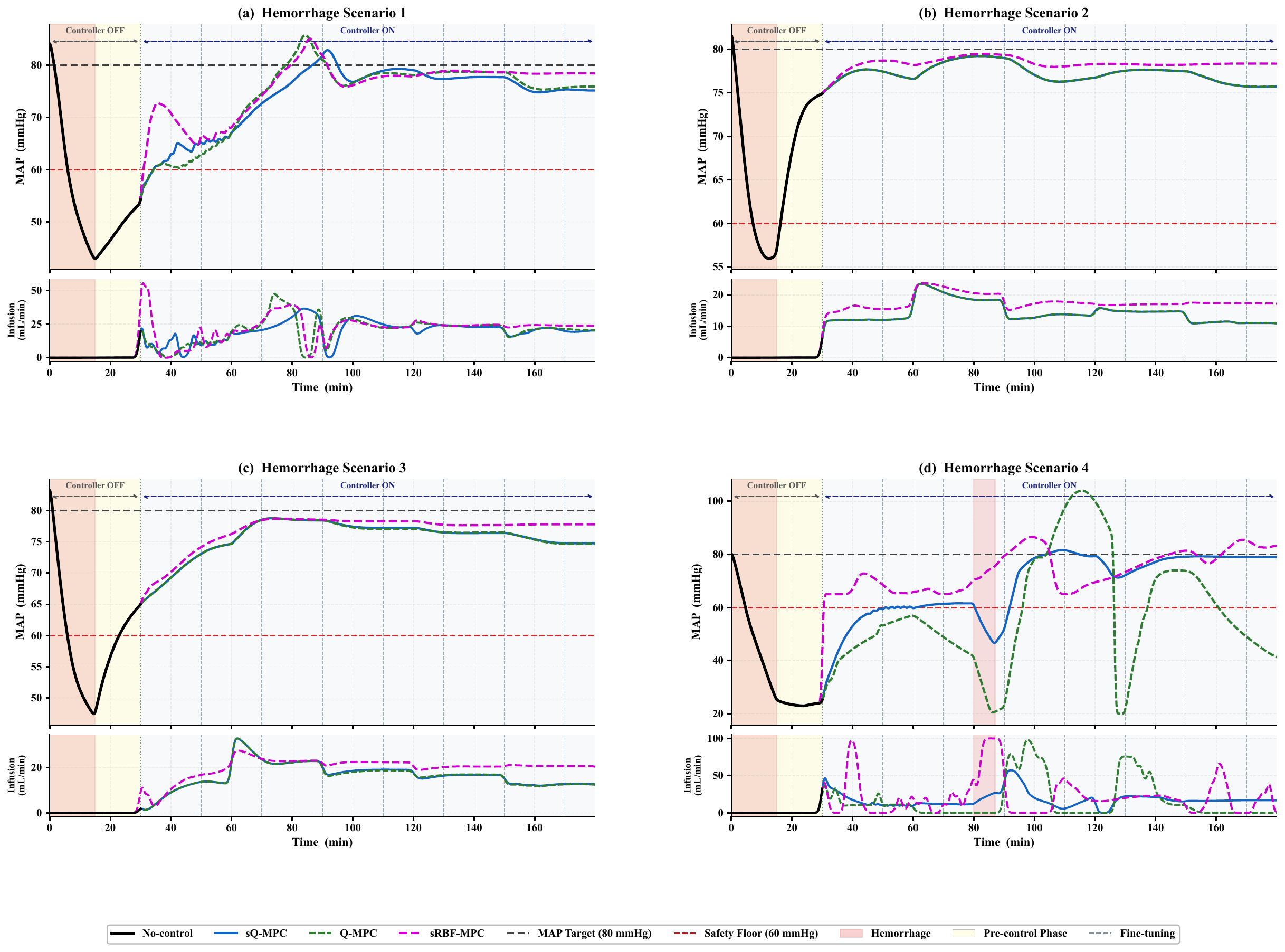}      
\caption{Closed-loop MAP regulation under four hemorrhage scenarios comparing Q-MPC, sQ-MPC, and sRBF-MPC. Each panel shows the patient MAP trajectory (top) and the corresponding infusion rate (bottom). The light-red bands mark the active hemorrhage window ($t = 0$--$15$ min); The light-yellow region marks the pre-control phase, during which no infusion is applied. The dotted vertical line at $t = 30$ min marks controller activation. Dashed purple vertical lines at $t = 60, 90, 120,$ and $150$ min indicate periodic fine-tuning events. The dashed gray and red horizontal lines indicate the MAP target (80 mmHg) and the safety floor (60 mmHg), respectively.} 
\label{fig:control_combined} 
\end{figure*}

Fig. ~\ref{fig:control_combined} compares the performance of the proposed sRBF-MPC against Q-MPC and sQ-MPC across four hemorrhage scenarios. 
The MAP and infusion trajectories reveal distinct control behaviors across the scenarios. In scenarios 2 and 3, all controllers successfully restored MAP and maintained values close to the MAP target of 80 mmHg. Because prediction uncertainty remained relatively small in these cases, the chance constraints were largely inactive, resulting in nearly identical MAP trajectories for the Q-MPC and sQ-MPC. The proposed sRBF-MPC achieved slightly improved tracking performance while maintaining smooth infusion profiles.

More pronounced differences emerged in scenarios 1 and 4, where larger disturbances activated the chance-constrained formulation. Closed-loop controller performance across hemorrhage scenarios are shown in Table~\ref{tab:control_results}. In scenario 1, the sRBF-MPC recovered from the hemorrhage-induced MAP drop more rapidly and reduced the time below the 60 mmHg to 0.6\%, compared with 2.8\% and 2.9\% for the sQ-MPC and Q-MPC, respectively. The advantage became more significant in scenario 4, where a secondary hemorrhage at $t = 80$~min induced a renewed MAP drop during the active control phase. The Q-MPC failed to maintain MAP above the safety floor for 64.2\% of the time after controller activation, while sQ-MPC improved this to 24.7\%. The proposed sRBF-MPC compensated for both the primary and secondary disturbances, keeping MAP below 60~mmHg for only 0.1\% of the time. This result demonstrates that the combination of uncertainty-aware prediction and chance-constrained optimization is particularly advantageous when recurrent hemorrhagic events occur mid-treatment. Across all four scenarios, the proposed sRBF-MPC consistently achieved the lowest performance error. In scenario 4, it reduced RMSE to 11.73\%, compared with 38.92\% and 21.04\% for Q-MPC and sQ-MPC, respectively, while also improving safety performance.

The infusion profiles highlight the trade-off introduced by uncertainty-aware control. The sRBF-MPC generally required higher infusion volumes than the deterministic MPC formulations, particularly in scenario 4. This behavior is a direct consequence of the chance constraints, which drive the controller toward more conservative actions when prediction uncertainty increases. By maintaining a larger safety margin above the lower MAP bound, the controller sacrifices some fluid efficiency in exchange for improved tracking accuracy and a reduced risk of hypotension.

The computational performance results reported in Table~\ref{tab:control_results} indicate that the proposed sRBF-MPC requires longer optimization times than the Q-MPC and sQ-MPC due to the combined effects of uncertainty propagation, chance-constraint evaluation, and RBF parameterization. Nevertheless, the average solve time remained below 300~ms per control update across all scenarios, which is substantially shorter than the 0.1-minute (6-second) sampling interval used in this study. These results demonstrate that the additional computational burden associated with uncertainty-aware control remains negligible relative to the system dynamics, supporting the feasibility of real-time implementation.

Overall, these results demonstrate that the proposed sRBF-MPC provides the most reliable MAP regulation under uncertainty, particularly in high-risk hemorrhage scenarios where safety becomes critical. The combination of uncertainty-aware prediction, chance-constrained optimization, smooth RBF parameterization, and online model adaptation enables robust closed-loop fluid resuscitation under physiologically challenging conditions.

  \begin{table*}[htbp]
  \centering
  \caption{Closed-Loop Performance Comparison of Q-MPC, sQ-MPC, and sRBF-MPC Across Four Hemorrhage Scenarios}
  \label{tab:control_results}
  \renewcommand{\arraystretch}{1.25}
  \begin{tabular}{clcccccr}
  \toprule
  \textbf{Scenario} & \textbf{Controller} & \textbf{RMSE (\%)} & \textbf{MAE (\%)} & \textbf{MDAPE (\%)}
    & \textbf{Below 60\,mmHg (\%)} & \textbf{Total Infusion (mL)} & \textbf{Solve Time/iteration (ms)} \\
  \midrule
  \multirow{3}{*}{1}
      & Q-MPC       & 11.09 & 7.61  & 4.58  & 2.9  & 3100.8 & $27.5  \pm 17.3$  \\
    & sQ-MPC      & 10.67 & 7.69  & 4.52  & 2.8  & 3011.8 & $62.3  \pm 65.7$  \\
    & sRBF-MPC & 8.13 & 5.67 & 2.57  & 0.6  & 3364.4 & $255.9 \pm 326.0$ \\
  \midrule
  \multirow{3}{*}{2}
      & Q-MPC       & 3.67 & 3.44 & 3.40 & 0.0 & 2145.8 & $14.0 \pm 6.9$  \\
    & sQ-MPC      & 3.68 & 3.45 & 3.40 & 0.0 & 2143.5 & $28.8 \pm 15.2$ \\

    & sRBF-MPC & 2.20 & 2.04 & 2.10 & 0.0 & 2663.8 & $55.4 \pm 31.3$ \\
  \midrule
  \multirow{3}{*}{3}
      & Q-MPC       & 6.65 & 5.54 & 4.43 & 0.0 & 2453.5 & $18.3 \pm 5.9$  \\
    & sQ-MPC      & 6.62 & 5.50 & 4.51 & 0.0 & 2468.5 & $37.4 \pm 13.0$ \\

    & sRBF-MPC & 5.35 & 4.01 & 2.78 & 0.0 & 3019.3 & $75.6 \pm 67.6$ \\
  \midrule
  \multirow{3}{*}{4}
    & Q-MPC      & 38.92 & 34.00 & 32.99 & 64.2 & 2614.1 & $25.6  \pm 13.4$ \\
    & sQ-MPC       & 21.04 & 14.52 & 7.01  & 24.7 & 2682.4 & $212.1 \pm 190.8$  \\
    & sRBF-MPC & 11.73 & 9.89 & 9.05  & 0.1 & 3348.8 & $265.6 \pm 373.1$ \\
  \bottomrule
  \end{tabular}

  \smallskip
  \end{table*}

\section{Discussion, Limitations, and Future Directions}\label{sec5}

This study introduced a holistic framework for automated fluid resuscitation that integrates novel patient modeling techniques with explicit uncertainty quantification and a computationally efficient closed-loop stochastic control strategy. In terms of modeling, the UVAE-SSM was introduced to represent hemodynamic dynamics while explicitly capturing aleatoric uncertainty from noisy and limited measurements. To further advance the framework, BNSSM was introduced, yielding a VPG capable of quantifying epistemic uncertainty and producing an ensemble of plausible patient trajectories. In terms of closed-loop control, a new discrete-time sRBF-MPC controller was introduced that propagates uncertainty to generate safety-aware optimal infusion policies for regulating MAP. By approximating nonlinear trajectories with smooth global basis functions, the sRBF-MPC controller achieved stable MAP regulation while maintaining probabilistic safety under uncertain dynamics, demonstrating the benefit of combining smooth control actions with uncertainty-aware modeling. Finally, an online adaptation scheme was incorporated, enabling the nominal model to be progressively fine-tuned during therapy. 

The proposed framework bridges the gap between data-driven physiological modeling and control-theoretic optimization, providing a cohesive approach that accounts for both inter- and intra-patient variability. Through explicit uncertainty quantification and online model personalization, the controller dynamically adapts to evolving patient conditions, while maintaining computational efficiency. Collectively, these features move the framework toward the broader goal of closed-loop resuscitation systems capable of learning, predicting, and adjusting therapy in real time to ensure both safety and efficacy in hemodynamic management.

Although the proposed automated resuscitation framework shows strong promise, its limitations should be carefully considered before drawing broader conclusions. First, the available animal dataset is relatively small for learning complex nonlinear physiological dynamics, which may limit the ability of the models to capture the full spectrum of patient variability. Both the nominal model and the Bayesian VPG were trained using animal data, which may limit generalizability to broader patient populations. To partially address this concern, the proposed models were additionally evaluated on an independent human clinical dataset and demonstrated consistent predictive performance across subjects. Furthermore, the treatment strategy was restricted to crystalloid infusion, whereas clinical resuscitation may involve multiple concurrent interventions such as blood product transfusion and vasopressor administration, whose physiological effects may depend on the composition of the intervention itself \cite{ghetmiri2021personalized}.

In light of the above limitations, future work will prioritize improving generalizability and clinical fidelity. Specifically, we will extend the predictive model beyond MAP to incorporate additional hemodynamic variables, enabling a more comprehensive characterization of patient state. Another key avenue for future work is to develop the VPG into a regulatory-grade testbed, analogous to the UVA/Padova simulator used in diabetes management \cite{cobelli2023developing, man2014uva}, to support robust preclinical evaluation of closed-loop resuscitation. Finally, the sRBF-MPC will be evaluated beyond simulation using a hardware-in-the-loop testbed \cite{mirinejad2019evaluation} that integrates computational models with medical devices such as an infusion pump and a patient monitor, to assess real-time feasibility and robustness under clinically realistic constraints.

\section{Conclusion}\label{sec6}
This study presented a data-driven framework for automated fluid resuscitation that integrates Bayesian physiological modeling and stochastic RBF optimal control. The UVAE-SSM provided a compact state-space representation with aleatoric uncertainty quantification, while the BNSSM enabled virtual patient generation through a Bayesian transition model. The proposed sRBF-MPC achieved stable MAP regulation, while demonstrating superior risk-aware performance compared to Q-SMPC and Q-MPC, and online adaptation improved personalization and robustness by accommodating evolving patient responses. These results demonstrate the potential of the proposed framework for uncertainty-aware, adaptive fluid management in critical care. Future work will extend the modeling scope to additional hemodynamic variables and evaluate the framework in hardware-in-the-loop and multi-therapy settings to enhance clinical relevance.


\begin{thebibliography}{10}
\providecommand{\url}[1]{#1}
\csname url@samestyle\endcsname
\providecommand{\newblock}{\relax}
\providecommand{\bibinfo}[2]{#2}
\providecommand{\BIBentrySTDinterwordspacing}{\spaceskip=0pt\relax}
\providecommand{\BIBentryALTinterwordstretchfactor}{4}
\providecommand{\BIBentryALTinterwordspacing}{\spaceskip=\fontdimen2\font plus
\BIBentryALTinterwordstretchfactor\fontdimen3\font minus \fontdimen4\font\relax}
\providecommand{\BIBforeignlanguage}[2]{{%
\expandafter\ifx\csname l@#1\endcsname\relax
\typeout{** WARNING: IEEEtran.bst: No hyphenation pattern has been}%
\typeout{** loaded for the language `#1'. Using the pattern for}%
\typeout{** the default language instead.}%
\else
\language=\csname l@#1\endcsname
\fi
#2}}
\providecommand{\BIBdecl}{\relax}
\BIBdecl

\bibitem{avital2022closed}
G.~Avital, E.~J. Snider, D.~Berard, S.~J. Vega, S.~I. Hernandez~Torres, V.~A. Convertino, J.~Salinas, and E.~N. Boice, ``Closed-loop controlled fluid administration systems: a comprehensive scoping review,'' \emph{Journal of personalized medicine}, vol.~12, no.~7, p. 1168, 2022.

\bibitem{mohammad2014educational}
F.~G. Bonanno, ``Management of hemorrhagic shock: physiology approach, timing and strategies,'' \emph{Journal of Clinical Medicine}, vol.~12, no.~1, p.~260, 2022.

\bibitem{parvinian2019credibility}
B.~Parvinian, P.~Pathmanathan, C.~Daluwatte, F.~Yaghouby, R.~A. Gray, S.~Weininger, T.~M. Morrison, and C.~G. Scully, ``Credibility evidence for computational patient models used in the development of physiological closed-loop controlled devices for critical care medicine,'' \emph{Frontiers in physiology}, vol.~10, p. 220, 2019.

\bibitem{estiri2023precision}
E.~Estiri and H.~Mirinejad, ``Precision dosing in critical care: application of machine learning in fluid therapy,'' in \emph{2023 IEEE International Conference on Digital Health (ICDH)}.\hskip 1em plus 0.5em minus 0.4em\relax IEEE, 2023, pp. 348--352.

\bibitem{bighamian2016lumped}
R.~Bighamian, A.~T. Reisner, and J.-O. Hahn, ``A lumped-parameter subject-specific model of blood volume response to fluid infusion,'' \emph{Frontiers in Physiology}, vol.~7, p. 390, 2016.

\bibitem{bighamian2018control}
R.~Bighamian, B.~Parvinian, C.~G. Scully, G.~Kramer, and J.-O. Hahn, ``Control-oriented physiological modeling of hemodynamic responses to blood volume perturbation,'' \emph{Control engineering practice}, vol.~73, pp. 149--160, 2018.

\bibitem{kim2014accuracy}
S.-H. Kim, M.~Lilot, K.~S. Sidhu, J.~Rinehart, Z.~Yu, C.~Canales, and M.~Cannesson, ``Accuracy and precision of continuous noninvasive arterial pressure monitoring compared with invasive arterial pressure: a systematic review and meta-analysis,'' \emph{Anesthesiology}, vol. 120, no.~5, pp. 1080--1097, 2014.

\bibitem{grant2023optimal}
J.~Grant and H.~Mirinejad, ``An optimal control approach for automated fluid resuscitation,'' \emph{arXiv preprint arXiv:2312.06521}, 2023.

\bibitem{jin2018development}
X.~Jin, R.~Bighamian, and J.-O. Hahn, ``Development and in silico evaluation of a model-based closed-loop fluid resuscitation control algorithm,'' \emph{IEEE Transactions on Biomedical Engineering}, vol.~66, no.~7, pp. 1905--1914, 2018.

\bibitem{meerza2021precise}
S.~I.~A. Meerza, A.~Affan, H.~Mirinejad, M.~E. Brier, J.~M. Zurada, and T.~Inanc, ``Precise warfarin management through personalized modeling and control with limited clinical data,'' in \emph{2021 43rd Annual International Conference of the IEEE Engineering in Medicine \& Biology Society (EMBC)}.\hskip 1em plus 0.5em minus 0.4em\relax IEEE, 2021, pp. 5035--5038.

\bibitem{alsalti2022design}
M.~Alsalti, A.~Tivay, X.~Jin, G.~C. Kramer, and J.-O. Hahn, ``Design and in silico evaluation of a closed-loop hemorrhage resuscitation algorithm with blood pressure as controlled variable,'' \emph{Journal of Dynamic Systems, Measurement, and Control}, vol. 144, no.~2, p. 021001, 2022.

\bibitem{estiri2024variational}
E.~Estiri and H.~Mirinejad, ``Variational autoencoder-based model predictive control for automated fluid resuscitation,'' in \emph{2024 IEEE EMBS International Conference on Biomedical and Health Informatics (BHI)}.\hskip 1em plus 0.5em minus 0.4em\relax IEEE, 2024, pp. 1--5.

\bibitem{mirinejad2015individualized}
H.~Mirinejad and T.~Inanc, ``Individualized anemia management using a radial basis function method,'' in \emph{2015 IEEE Great Lakes Biomedical Conference (GLBC)}.\hskip 1em plus 0.5em minus 0.4em\relax IEEE, 2015, pp. 1--4.

\bibitem{kao2025development}
Y.-M. Kao, Y.~R. Chalumuri, C.~M. Sampson, S.~A. Shah, J.~R. Salsbury, A.~Tivay, M.~Kinsky, G.~C. Kramer, and J.-O. Hahn, ``Development of a virtual patient generator for simulation of vasopressor resuscitation,'' \emph{Journal of Dynamic Systems, Measurement, and Control}, vol. 147, no.~3, p. 031001, 2025.

\bibitem{cobelli2023developing}
C.~Cobelli and B.~Kovatchev, ``Developing the uva/padova type 1 diabetes simulator: modeling, validation, refinements, and utility,'' \emph{Journal of Diabetes Science and Technology}, vol.~17, no.~6, pp. 1493--1505, 2023.

\bibitem{man2014uva}
C.~D. Man, F.~Micheletto, D.~Lv, M.~Breton, B.~Kovatchev, and C.~Cobelli, ``The uva/padova type 1 diabetes simulator: new features,'' \emph{Journal of diabetes science and technology}, vol.~8, no.~1, pp. 26--34, 2014.

\bibitem{coulibaly2023personalized}
L.~M. Coulibaly, S.~Sacu, P.~Fuchs, H.~Bogunovic, G.~Faustmann, C.~Unterrainer, G.~S. Reiter, and U.~Schmidt-Erfurth, ``Personalized treatment supported by automated quantitative fluid analysis in active neovascular age-related macular degeneration (namd)—a phase iii, prospective, multicentre, randomized study: design and methods,'' \emph{Eye}, vol.~37, no.~7, pp. 1464--1469, 2023.

\bibitem{mirinejad2019evaluation}
H.~Mirinejad, B.~Parvinian, M.~Ricks, Y.~Zhang, S.~Weininger, J.-O. Hahn, and C.~G. Scully, ``Evaluation of fluid resuscitation control algorithms via a hardware-in-the-loop test bed,'' \emph{IEEE Transactions on Biomedical Engineering}, vol.~67, no.~2, pp. 471--481, 2019.

\bibitem{rafie2004hypotensive}
A.~D. Rafie, P.~A. Rath, M.~W. Michell, R.~A. Kirschner, D.~J. Deyo, D.~S. Prough, J.~J. Grady, and G.~C. Kramer, ``Hypotensive resuscitation of multiple hemorrhages using crystalloid and colloids,'' \emph{Shock}, vol.~22, no.~3, pp. 262--269, 2004.

\bibitem{estiri2023closed}
E.~Estiri and H.~Mirinejad, ``Closed-loop control of fluid resuscitation using reinforcement learning,'' \emph{IEEE Access}, vol.~11, pp. 140\,569--140\,581, 2023.

\bibitem{pinsky2024autonomous}
M.~R. Pinsky, H.~Gomez, F.~X. Guyette, L.~Weiss, A.~Dubrawski, J.~Leonard, R.~MacLachlan, L.~Gordon, T.~Lagattuta, D.~Salcido \emph{et~al.}, ``Autonomous precision resuscitation during ground and air transport of an animal hemorrhagic shock model,'' \emph{Intensive Care Medicine Experimental}, vol.~12, no.~1, p.~44, 2024.

\bibitem{ganapathy2023precision}
A.~S. Ganapathy, N.~T. Patel, A.~P. Wiley, M.~R. Lane, J.~E. Jordan, M.~A. Johnson, J.~Y. Adams, L.~P. Neff, and T.~K. Williams, ``Precision automated critical care management: Closed-loop critical care for the treatment of distributive shock in a swine model of ischemia-reperfusion,'' \emph{Journal of Trauma and Acute Care Surgery}, vol.~95, no.~4, pp. 490--496, 2023.

\bibitem{mirinejad2021radial}
H.~Mirinejad, T.~Inanc, and J.~M. Zurada, ``Radial basis function interpolation and galerkin projection for direct trajectory optimization and costate estimation,'' \emph{IEEE/CAA Journal of Automatica Sinica}, vol.~8, no.~8, pp. 1380--1388, 2021.

\bibitem{mirinejad2017individualized}
H.~Mirinejad, A.~E. Gaweda, M.~E. Brier, J.~M. Zurada, and T.~Inanc, ``Individualized drug dosing using rbf-galerkin method: Case of anemia management in chronic kidney disease,'' \emph{Computer methods and programs in biomedicine}, vol. 148, pp. 45--53, 2017.

\bibitem{estiri2026automated}
E.~Estiri and H.~Mirinejad, ``Automated fluid resuscitation via robust nonlinear state-space modeling and radial basis function optimal control,'' \emph{Journal of Dynamic Systems, Measurement, and Control}, pp. 1--30, 2026.

\bibitem{bowman2016generating}
S.~Bowman, L.~Vilnis, O.~Vinyals, A.~Dai, R.~Jozefowicz, and S.~Bengio, ``Generating sentences from a continuous space,'' in \emph{Proceedings of the 20th SIGNLL conference on computational natural language learning}, 2016, pp. 10--21.

\bibitem{fu2019cyclical}
H.~Fu, C.~Li, X.~Liu, J.~Gao, A.~Celikyilmaz, and L.~Carin, ``Cyclical annealing schedule: A simple approach to mitigating kl vanishing,'' in \emph{Proceedings of the 2019 Conference of the North American Chapter of the Association for Computational Linguistics: Human Language Technologies, Volume 1 (Long and Short Papers)}, 2019, pp. 240--250.

\bibitem{kendall2017uncertainties}
A.~Kendall and Y.~Gal, ``What uncertainties do we need in bayesian deep learning for computer vision?'' \emph{arXiv preprint arXiv:1703.04977}, 2017.

\bibitem{blundell2015weight}
C.~Blundell, J.~Cornebise, K.~Kavukcuoglu, and D.~Wierstra, ``Weight uncertainty in neural network,'' in \emph{International conference on machine learning}.\hskip 1em plus 0.5em minus 0.4em\relax PMLR, 2015, pp. 1613--1622.

\bibitem{mirinejad2015rbf}
H.~Mirinejad and T.~Inanc, ``Rbf method for optimal control of drug administration in the anemia of hemodialysis patients,'' in \emph{2015 41st Annual Northeast Biomedical Engineering Conference (NEBEC)}.\hskip 1em plus 0.5em minus 0.4em\relax IEEE, 2015, pp. 1--2.

\bibitem{mirinejad2015radial}
H.~Mirinejad and T.~Inanc, ``A radial basis function method for direct trajectory optimization,'' in \emph{2015 American Control Conference (ACC)}.\hskip 1em plus 0.5em minus 0.4em\relax IEEE, 2015, pp. 4923--4928.

\bibitem{drobin1999volume}
D.~Drobin and R.~G. Hahn, ``Volume kinetics of ringer's solution in hypovolemic volunteers,'' \emph{Anesthesiology}, vol.~90, no.~1, pp. 81--91, 1999.

\bibitem{ghetmiri2021personalized}
D.~E. Ghetmiri, M.~J. Cohen, and A.~A. Menezes, ``Personalized modulation of coagulation factors using a thrombin dynamics model to treat trauma-induced coagulopathy,'' \emph{npj Systems Biology and Applications}, vol.~7, no.~1, p.~44, 2021.

\end{thebibliography}

\begin{IEEEbiography}[{\includegraphics[width=1.05in,height=1.25in,clip,keepaspectratio]{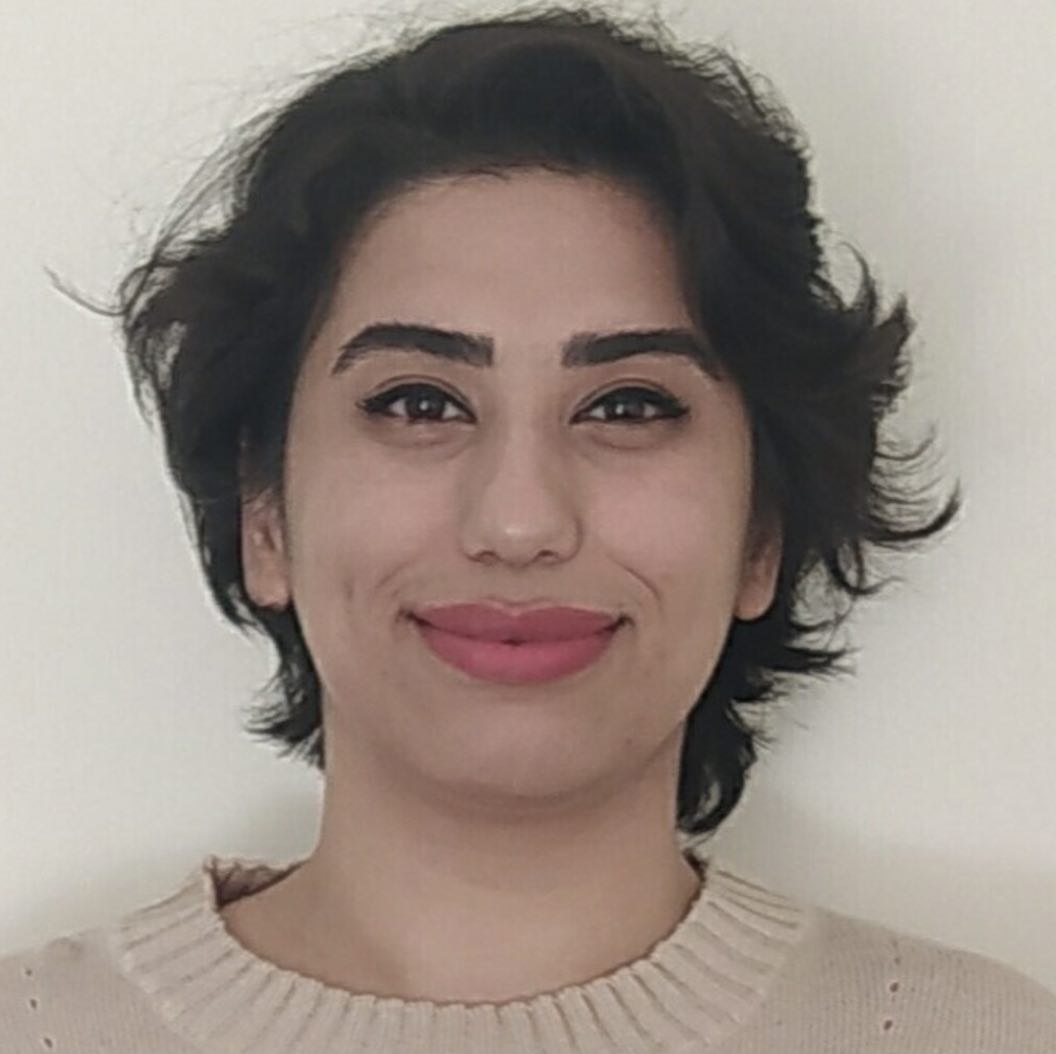}}]{Elham Estiri} (Student Member, IEEE) received her B.S. degree in electrical engineering from the Hamedan University of Technology, and her M.S. degree in biomedical engineering from the Hakim Sabzevari University. She is currently  a Graduate Research Assistant with the School of Engineering, College of Aeronautics and Engineering, Kent State University, Kent, OH, USA. Her current research is concerned with the design, modeling, control, and testing of autonomous algorithms for medication dosing. 
\end{IEEEbiography}
\begin{IEEEbiography}[{\includegraphics[width=1.25in,height=1.25in,clip,keepaspectratio]{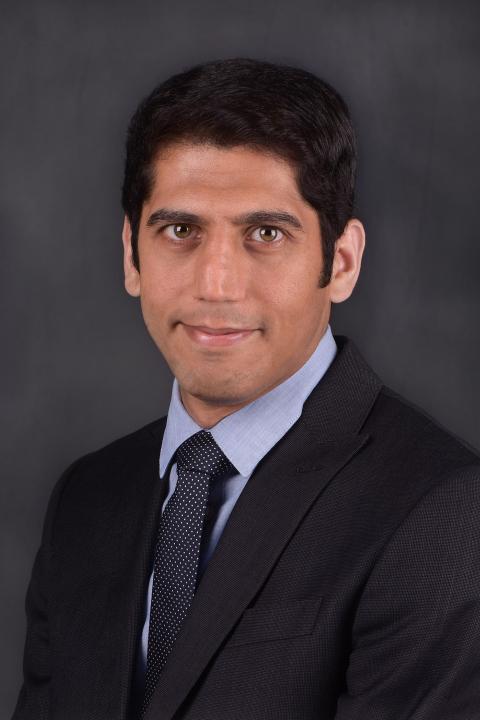}}]{Hossein Mirinejad}  (Senior Member, IEEE) received
the Ph.D. degree in electrical engineering from
the University of Louisville, Louisville, KY, USA,
in 2016. He is currently an Associate Professor
with the School of Engineering, College of Aeronautics and Engineering,
Kent State University, Kent, OH, USA. Prior to
joining Kent State, he was a Research Fellow
with the Food and Drug Administration (FDA) from 2017 to 2019, and a Postdoctoral
Fellow with the University of Michigan from 2016 to 2017. His research focuses on advancing the modeling, control, and testing of autonomous systems using principles of dynamical systems, machine learning, and automatic control theory. 
\end{IEEEbiography}

\vfill

\end{document}